\documentclass[
 reprint,
superscriptaddress,
]{revtex4-2}
\usepackage{amsmath} 
\usepackage{graphicx}
\usepackage{dcolumn}
\usepackage{bm}
\begin{document}

\preprint{APS/123-QED}
\title{Electronic and magnetic ground state of 4$d^3$ double perovskite ruthenates A$_2$LaRuO$_6$ (A $=$ Ca, Sr, Ba)}

\author{Asha Ann Abraham}
 
 \affiliation{%
Department of Physics, Indian Institute of Technology Palakkad, Kerala-678623, India\\
}%
\author{Roumita Roy}
\affiliation{%
 School of Physical Sciences,
Indian Institute of Technology, Goa, Farmagudi, Ponda Goa 403401, India \\
}%
\author{Ruta Kulkarni}
\affiliation{%
Department of Condensed Matter Physics and Materials Science, Tata Institute of
Fundamental Research, Homi Bhabha Road, Colaba, Mumbai 400005, India\\
}%
\author{Sudipta Kanungo}
\affiliation{%
 School of Physical Sciences,
Indian Institute of Technology, Goa, Farmagudi, Ponda Goa 403401, India \\
}%
\author{Soham Manni}%
\affiliation{%
Department of Physics, Indian Institute of Technology Palakkad, Kerala-678623, India\\
}%
\thanks{\email{smanni@iitpkd.ac.in}}
\date{\today}
\begin{abstract}
4$d$ transition metal oxide (TMO) offers an intriguing puzzle for their electronic and magnetic ground state. They are in the cross-over regime of strong spin orbit interaction (SOI) and electron-electron correlation ($U$) with quenched orbital angular momentum. Our work unravels the electronic and magnetic ground state of the less investigated 4$d^{3}$  double perovskite ruthenates A$_{2}$LaRuO$_6$ (A = Ca, Ba). The negligible effect of SOI is evident from the bulk magnetic, specific heat measurements and density functional theory (DFT) calculations, indicating a classical spin-only magnetic ground state (${S}$ = 3/2) for the materials. Magnetization measurements show that both materials have long range antiferromagnetic order with high degree of magnetic frustration ($f$ $\approx$13 -15). Interestingly, a near $T^2$- behavior is observed in low-$T$ magnetic heat capacity measurement, indicating the presence of low-dimensional spin-wave exciation and magnetic frustration in both materials. The temperature dependent resistivity measurements and electronic band structure calculations confirm a conventional Mott insulating ground state in these two systems. Moreover, our experimental investigation and DFT calculations highlight the reason for the nonexistence of Sr$_2$LaRuO$_6$. 
\end{abstract}
\maketitle


\section{\label{sec:level1}Introduction}

Spin-orbit coupling (SOC) is a crucial energy scale that often determines the electronic ground state in 4$d$ and 5$d$ transition metal oxides (TMOs). By varying the relative strength of SOC and on-site electron-electron interaction ($U$), a multitude of exotic quantum phases of matter are predicted in 4d/5d TMOs, e.g., topological insulators, spin-orbit coupled Mott insulators, axion insulators, Weyl semimetals, quantum spin liquids, etc.\cite{annual_condensed_matter}. The discovery of the unconventional spin-orbt coupled Mott insulating state in Sr$_2$IrO$_4$ ($J_{eff}$ = 1/2) and its similarity to cuprates created renewed interest in 4d/5d TMOs, especially with perovskite structure\cite{Kim_novel,SIO_dwavegap,SIO_pseudogap,SIO_pseudogap_STM,SIO_fermiarc_ARPEs}. In 4d TMOs, SOC and $U$ are comparable and often compete with each other\cite{energyscales}. They display an amalgam of phases ranging from the correlated metallic phase in Sr$_2$RhO$_4$, Mott insulating states in Ca$_2$RuO$_4$ , heavy Fermi liquid behavior in CaRuO$_3$ to superconductivity in Sr$_2$RuO$_4$ \cite{Sr_2RhO_4, Ca2RuO4, CaRuO3, Sr2RuO4}. Often they present a highly tunable fragile ground state, which can be perturbed by external perturbations such as charge carrier doping and pressure \cite{Ca2RuO4_doping_2000, SrdopedLa2CuO4_2021, pressure_phasetransition_LaMnO3_2015,LaMnO3_DFT}.

Double Perovskite (DP) TMOs serve as an ideal platform for studying tunability of 4d-TMOs that contain different atomic sites for magnetic and nonmagnetic atoms. In DPs (A$_2$BB$^{'}$O$_{6}$), the A site is usually occupied by an alkaline/ alkaline earth metal and B$/$B$^{'}$-site can host an alkaline/ alkaline earth metal or a transition metal. In a mixed DP TMO system where 3d and 5d magnetic TM ions occupy the B and B$^{'}$ site, respectively, the hybridization of 3d-5d orbitals controls the electronic and magnetic properties e.g. La$_2$CoIrO$_{6}$, La$_2$CoPtO$_{6}$\cite{3d5d_LCIO}. Therefore, it is difficult to study the effect of pure SOC in mixed DPs. On the other hand, ordered DPs with a single magnetic 4d/5d TM ion at the B$^{\prime}$ site and B site by alkaline / alkaline earth / non-magnetic rare-earth elements are the ideal candidates for studying the competition between SOC and $U$,  which can also be tuned by systematic chemical doping at the A/B site. Many novel ground states are realized in these materials, e.g. octupolar ordering in Ba$_2$MgOsO$_6$, quadrupolar ordering in Ba$_2$MgReO$_6$, valence bond glass state in Ba$_2$YMoO$_6$, magentically ordered state with Kitaev interactions in La$_2$ZnIrO$_6$ and La$_2$MgIrO$_6$~\cite{octupolarordering_Ba2MgOsO6, Ba2MgReO6_Quadrupolar, Valencebondstate_BYMO, LZIO_LMIO_Kitaev}.

Among these DPs, 4d$^{3}$/5d$^{3}$-TMO remains an outlier, where a classical $\textit{S}$ = 3/2 ground state is predicted due to the total quenching of the orbital angular momentum ($\textit{L}$ = 0) of the t$_{2g}$ orbital by octahedral crystal electric field (CEF)\cite{classicalspin}. In contrast to the predicted classical spin-only ground state, some of the 5d$^{3}$-DP TMO systems show a relativistic spin-orbit coupled $J_{eff}$ = 3/2 ground state e.g. Ba$_2$YOsO$_6$,  Sr$_2$MgIrO$_6$ and Sr$_2$CaIrO$_6$ \cite{Ba2YOsO6,Sr2MgIrO6_Sr2CaIrO6}. 4d- TMOs, specifically perovskite ruthenates, exhibit exotic ground states ranging from superconductivity to quantum criticality due to comparable $U$ and SOC \cite{CaRuO3_heavyfermiliquid, SrRuO3_Weylfermions, Sr2RuO4_superconductivity}. 4d$^{3}$-DP ruthenates with a single magnetic Ru ion can be an ideal platform to study the competition between different energy scales and its tunability. Recent investigations in the 4d$^3$ system, Ba$_2$YRuO$_6$ indicates the presence of unconventional multipolar ordering in it\cite{BYRO_multipolar}. In addition, the ambiguity that prevails around the superconducting nature of Cu-doped Sr$_2$YRuO$_6$ creates an additional interest in this class of ordered DP-ruthenates\cite{CudopedSr2YRuO6}.

Among A$_2$BRuO$_6$ (A =Ca, Sr, Ba ; B = Sc, Y, La) systems, A$_2$LaRuO$_6$ lacks a deeper understanding of its electronic and magnetic ground state. There are also no investigations to understand the effect of SOC on the electronic and magnetic ground state of the A$_2$BRuO$_6$. In this work, we have investigated the electronic and magnetic ground state of the A$_2$LaRuO$_6$ system in detail by bulk magnetic, specific heat, resistivity measurements and DFT calculations considering all the competing energy scales. Our results show a successful synthesis of polycrystalline Ca$_2$LaRuO$_6$ (CLRO) and Ba$_2$LaRuO$_6$ (BLRO) and a structural instability in the formation of Sr$_2$LaRuO$_6$ (SLRO). We studied the structural instability experimentally through Sr-doping in CLRO and BLRO along with the calculation of the phonon density of states for SLRO. Magnetization, specific heat, resistivity measurements, and density functional theory (DFT) calculations reveal the occurrence of an antiferromagnetic (AFM) Mott insulating ground state in A$_2$LaRuO$_6$ (A = Ca,Ba) system rather than a spin-orbit coupled Mott insulating state.
\setlength{\parskip}{0pt}
\section{\label{sec:level2}Experimental Details}
Polycrystalline A$_2$LaRuO$_6$ (A = Ca, Ba) was synthesized by the conventional solid-state synthesis route using stoichiometric quantities of ACO$_3$ (A = Ca, Ba), Ru powder and La$_2$O$_3$ (all with purity $ \geq $ 99.9\%). A homogeneous mixture of precursors was heated at 1000$^{\circ}C$, 1100$^{\circ}C$ and 1200$^{\circ}C$ in a high-temperature programmable Muffle furnace, along with intermediate grinding and pelletization. At each step, the samples were annealed for 24 hours. We obtained single-phase BLRO and CLRO polycrystalline powder. We tried to synthesize SLRO polycrystalline powder in the same route but failed to obtain the single-phase SLRO. The result was the same even after the temperature profile and initial precursors were modified. To understand the structural stability of SLRO, the A site of the A$_2$LaRuO$_6$ (A = Ca, Ba) was doped with 25$\%$ 50$\%$ and 75$\%$ Sr using the same synthesis route.

The structural characterization of the synthesized samples was carried out using powder X-ray diffraction (PXRD) at room temperature using a Rigaku Smart Lab X-ray diffractometer with Cu-K$_\alpha$ radiation ($\lambda$ = 1.54 \AA). The PXRD data was fitted using General Structural and Analysis Software II (GSAS II)by the Rietveld refinement method\cite{gsas}. The magnetization measurements were performed using the Quantum Design Magnetic Property Measurement System (MPMS) in the temperature range of 1.8 K - 300 K and magnetic field range between 0 - 70 kOe. A Quantum Design Physical Properties Measurement System (PPMS) was used for heat capacity measurements in the temperature range of 2 -300 K at 0 kOe and 90 kOe magnetic field. Resistivity measurements were carried out on cuboidal-shaped highly pressurized pellets using Quazar Tech's XPLORE 1.2 Physical Quantities Measurement System (PQMS) in the temperature range of 80 - 300 K at zero field by applying a constant voltage of 1V.
\section{Calculation methodology}
We performed the DFT calculations  within the plane-wave based basis set of the pseudopotential framework with Perdew-Burke-Ernzerhof (PBE)\cite{perdew1996generalized} exchange-correlation functional as implemented in the Vienna ${ab-initio}$ simulation package (VASP)\cite{kresse2001vasp,PhysRevB.54.11169}. The effect of electron-electron Coulomb correlations was taken into  account via onsite Hubbard $U$ \cite{anisimov1993density,dudarev1998electron}. The $U_{eff}$ ($U$$-$$J_H$) for the Ru-4$\textit{d}$ states was taken as 4 eV.  The SOC effect has been incorporated in the calculations through relativistic corrections to the original Hamiltonian \cite{hobbs2000fully}. The energy cut-off was set at 400 eV for both BLRO and CLRO. For the self-consistent calculations, we utilised a 6$\times6\times6$ and 6$\times6\times4$ k-mesh in the Brillouin zone (BZ) for cubic BLRO and monoclinic CLRO respectively. The experimentally obtained structures were optimized by relaxing the atomic positions towards equilibrium until the Hellmann-Feynman force becomes less than 0.001 eV/$\AA$, with the lattice parameters fixed at the experimentally obtained values. The dynamical stability of the crystal structures was investigated via phonon DOS, using the open-source PHONOPY code \cite{togo2015first}. 
\section{\label{sec:citeref}{Results and Discussions}}
\subsection{Structural Characterization}
\begin{figure}[h]
\centering
\includegraphics[width=\columnwidth]{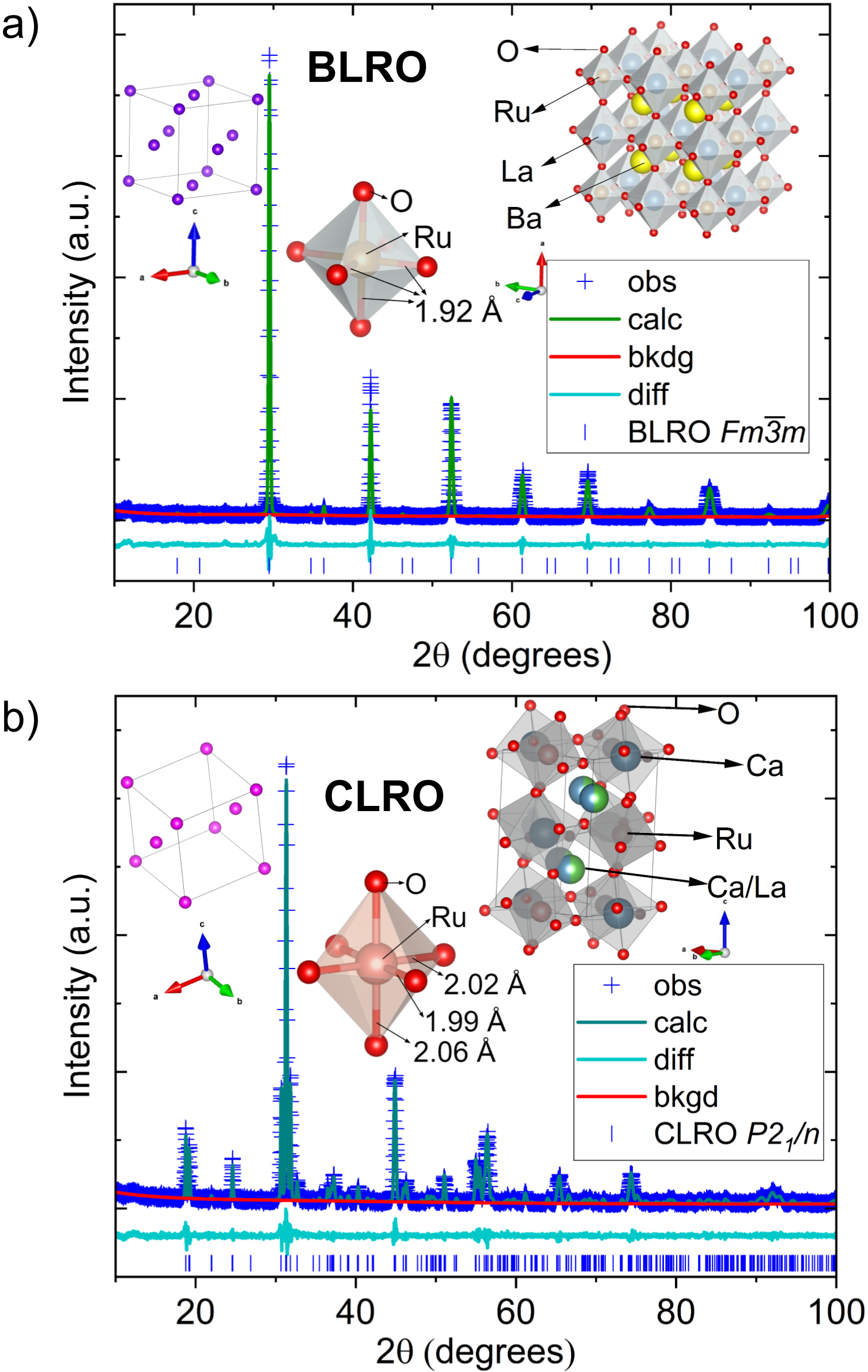}
\caption{Reitveld refinement of PXRD pattern a) BLRO and b) CLRO. The measured data, the calculated pattern from the refinement, the background and the difference between the measured data and the calculated pattern are referred to as obs, calc, bkgd, and diff, respectively. The vertical blue lines refer to the location of Bragg peaks for the respective phases. The crystal structure of BLRO and CLRO, their RuO$_6$ octahedron with Ru-O bond lengths and the corresponding Ru sublattice are shown in the respective insets, which are visualized using the VESTA software\cite{vesta}.}
\label{rr}
\end{figure}
From the PXRD measurements, the single-phase nature of the synthesized BLRO and CLRO was confirmed without any detectable impurity. Rietveld refinement for BLRO is done with the space group \textit{Fm$\bar{3}$m} (Space group no : 225) (FIG. \ref{rr}a). The same space group was previously reported for this material and for Ba$_{2}$YRuO$_{6}$ \cite{BLROFm3m, BYROFm3m}. 
We noticed that \textit{Fm$\bar{3}$m} provided a better fit for the synthesized BLRO compared to other reported structure \textit{R$\bar{3}$} \cite{BLRO_R3bar}. For the CLRO system, the Rietveld refinement was done using space group \textit{P2$_1$/n} (Space group no : 14) (FIG. \ref{rr}b) which was consistent with the previous report \cite{CLROP21byn}. TABLE \ref{RRlatticeparameters} provides the refined lattice parameters of A$_2$LaRuO$_6$ (A = Ca, Ba) obtained from their respective refinements. BLRO shows an ordered DP structure without any structural distortion that has a perfect RuO$_6$ octahedra with a single Ru-O bond length of 1.92 \AA (FIG. \ref{rr}a). Ru sublattice forms an FCC unit cell (FIG. \ref{rr}a). In CLRO, a significant distortion is observed, which forms a distorted RuO$_6$ octahedra with bond lengths of 2.06 \AA, 1.99 \AA~and 2.02 \AA~(FIG. \ref{rr}b). A different Ru sublattice is formed in CLRO as shown in the inset of FIG. \ref{rr}b. Single-phase PXRD patterns were obtained for 25\% and 50\% Sr-doped CLRO which are refined with \textit{P2$_1$/n} space group, the same as the pure CLRO (FIG.\ref{clro_u_d}).  However, multi-phase sample was obtained during the synthesis of Ca$_{0.5}$Sr$_{1.5}$LaRuO$_6$ (75\% Sr-doped CLRO) and all Sr-doped BLRO. The refined lattice parameters of Sr-doped CLRO are listed in Table \ref{rrdoped} in Appendix \ref{appendix}. A linear increase in the lattice volume was observed in Ca$_{1-x}$Sr$_{x}$LaRuO$_6$ (x = 0, 0.5, 1) series (Appendix \ref{appendix}, FIG. \ref{clro_u_d}).
\begin{table}
\caption{Refined lattice parameters of CLRO and BLRO. Here, R$_w$ represents the Reitveld refinement quality factor.}
\vspace{5 mm}
\centering
\begin{tabular}{ccccccc}
\hline
&& BLRO &&& CLRO \\ \hline
Crystal System&& Cubic&&& Monclinic  \\
Space group&& \textit{Fm$\bar{3}$m}  &&& \textit{P2$_1$/n}  \\
a ($\AA$)&& 8.5427(1)&&& 5.6127(1)\\
b ($\AA$)&& 8.5427(1)&&& 5.816(9)\\
c ($\AA$)&& 8.5427(1)&&& 8.0538(2)\\
$\beta$ ($^{\circ}$)&& 90.00(3)&&& 90.224(3)\\
R$_w$&&14.89 \%  &&&13.61 \%  \\ \hline
\end{tabular}
\label{RRlatticeparameters}
\end{table}
\begin{figure}[]
        \centering
         \includegraphics[width=0.98\columnwidth]{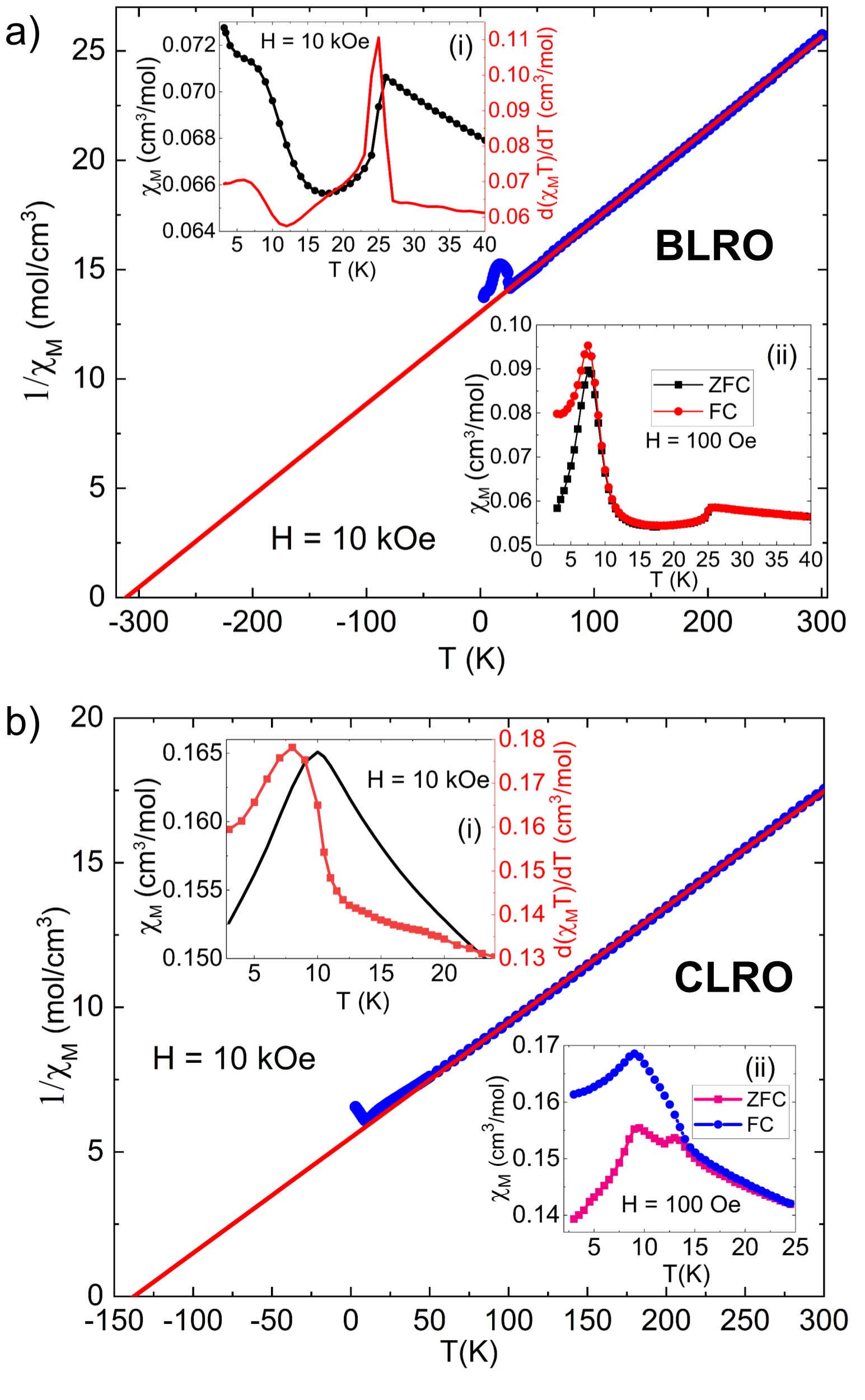}
         \caption{Temperature dependent inverse magnetic molar susceptibility data for a) BLRO  and b) CLRO at an external magnetic field, $H$ = 10 kOe with the CW fit (red line). Inset (i) shows the molar susceptibility ($\chi_M$) and $\frac{\mathrm{d}(\chi_{M}T)}{\mathrm{d} T}$ $vs.$ $T$ and inset (ii) shows the zero field cooled (ZFC) and field cooled (FC) $\chi_M$ $vs.$ $T$ at low field, $H$ = 100 Oe near magnetic phase transitions for both the systems in their respective panels.}
         \label{MT_u}
\end{figure}
\subsection{Magnetization Measurements}
FIG.  \ref{MT_u} shows the temperature dependent inverse magnetic molar susceptibility ($1/\chi_M = H/M$) of BLRO  and CLRO, measured in an external magnetic field, H = 10 kOe. Both follow a Curie-Weiss (CW) behavior at high temperature, $1/\chi_M = (T - \Theta)/C$. Here, $C$ is the Curie constant that provides the effective magnetic moment, $\mu_{eff}$ (= $\sqrt{8C} \mu_{B}$) and $\Theta$ is the CW temperature. The CW fit in the temperature range of 50 K to 300 K provides an effective magnetic moment, $\mu_{eff}$ = 3.907(2) $\mu_{B}$ for BLRO and 4.004(4) $\mu_{B}$ for CLRO, respectively. A slight variation in $\mu_{eff}$ ($<$ 1\% ) and $\Theta$ ($<$~ 5 \%) are observed for the CW fit in different temperature ranges between 50 K and 300 K. The experimental $\mu_{eff}$s are very similar to the theoretical spin-only value of $\mu_{eff}$ (= 3.87 $\mu_{B}$) for S = 3/2 in the 4d$^3$ system \cite{A2ScRuO6,magmoment_1,magmoment_2}. This indicates a negligible effect of SOC in the A$_2$LaRuO$_6$ systems. A large negative $\Theta$ of - 311.4(6) K and -137.5(3) K is obtained from the CW fit for BLRO and CLRO, respectively, which implies a strong antiferromagnetic exchange interaction between the Ru$^{5+}$ ions in both systems.

We have analyzed the magnetic ground state from the low-temperature magnetic susceptibility data. In the case of BLRO, the high field (H = 10 kOe) $\chi_{M}$ drops sharply below 25 K, which increases again at low temperature with a weak anomaly around 7 K. The transitions are prominent in $\frac{\mathrm{d}(\chi_{M}T)}{\mathrm{d} T}$ $vs.$~$T$ data (FIG \ref{MT_u}a (i)). The temperature dependent field cooled (FC) and zero field cooled (ZFC) $\chi_{M}$ measured at a lower field (H = 100 Oe) also shows a kink near 25~K without any hysteresis (FIG. \ref{MT_u}a(ii)). However, a small hysteresis is observed near 7~K. This behavior suggests the appearance of an antiferromagnetic ordering below $T_N$ = 25 K and spin canting below 7 K in BLRO. For CLRO, a sharp drop is observed around $T$ = 9 K in the high field ($H$ = 10 kOe) $\chi_{M}$ and $\frac{\mathrm{d}(\chi_{M}T)}{\mathrm{d} T}$ data (FIG.  \ref{MT_u}b(i)). The anomaly is also present in the low field ($H$ = 100 Oe) temperature dependent ZFC - FC $\chi_{M}$ data  (FIG. \ref{MT_u}b(ii)). Furthermore, strong hysteresis with a kink at around 13 K is observed in the 100 Oe, ZFC - FC $\chi_{m}$ data, above $T_N$. This additional transition is absent in the high-field $\chi_{m}$ data. It seems that upon lowering the temperature, spin freezing occurs in CLRO below 13 K and Ru$^{5+}$ spins orders antiferromagnetically below 9 K upon further cooling.  Both BLRO and CLRO have an antiferromagnetic ground state with a high frustration index ($f = |\Theta|/T_{N})$) 12.46 and 15.52 respectively. The magnetic frustration in BLRO originates from FCC Ru sublattice and the same in CLRO from isosceles triangles in the Ru sublattice (insets of FIG. \ref{rr}). The temperature dependent $\chi_M$ for Ca$_{1-x}$Sr$_x$LaRuO$_6$ (x = 0.5,1) is shown in the Appendix (FIG. \ref{MT_d}). The corresponding transition temperatures ($T_N$) ,$\Theta$ and $f$ are listed in Appendix, TABLE. \ref{MT_utable}.
\begin{figure}[ht]
         \centering
         \includegraphics[width=\columnwidth]{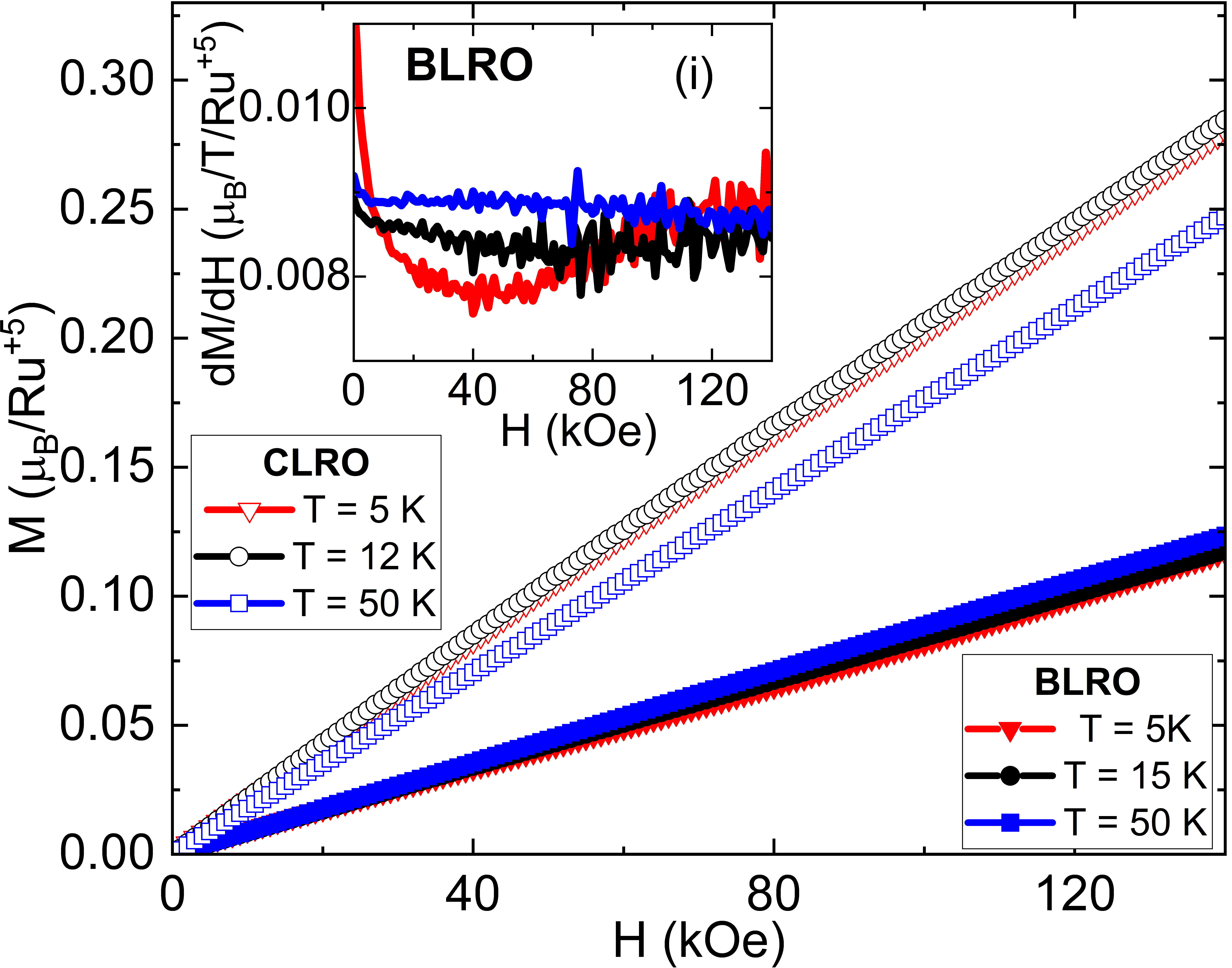}
         \caption{Field dependent magnetization ($M$ $vs.$ $H$) data for BLRO (closed symbols
         ) and CLRO (open symbols) at different temperatures. Inset (i) shows the $\frac{\mathrm{d} M}{\mathrm{d} H} $ $vs.$ $H$ plot of BLRO for the same temperatures.}
         \label{MH_u}
     \end{figure}
Field-dependent magnetization isotherms at different temperatures for BLRO and CLRO are shown in FIG. \ref{MH_u}. The AFM ground state is clearly evident from the linear $M~vs.~H~$ plot for BLRO at 15 K and 50 K and for CLRO at 5 K, 12 K and 50 K. However, for BLRO, a clear non-linearity is observed in $M~vs.~H~$ at low temperature ($T$ = 5K) which is clearly reflected in $\frac{\mathrm{d}M}{\mathrm{d} H} $ $vs.$~$H$ plot (FIG. \ref{MH_u}a(i)). This might be due to the canted AFM ground state in this system.

\subsection{Resistivity Measurements}
\begin{figure}[]
         \centering
         \includegraphics[width=\columnwidth]{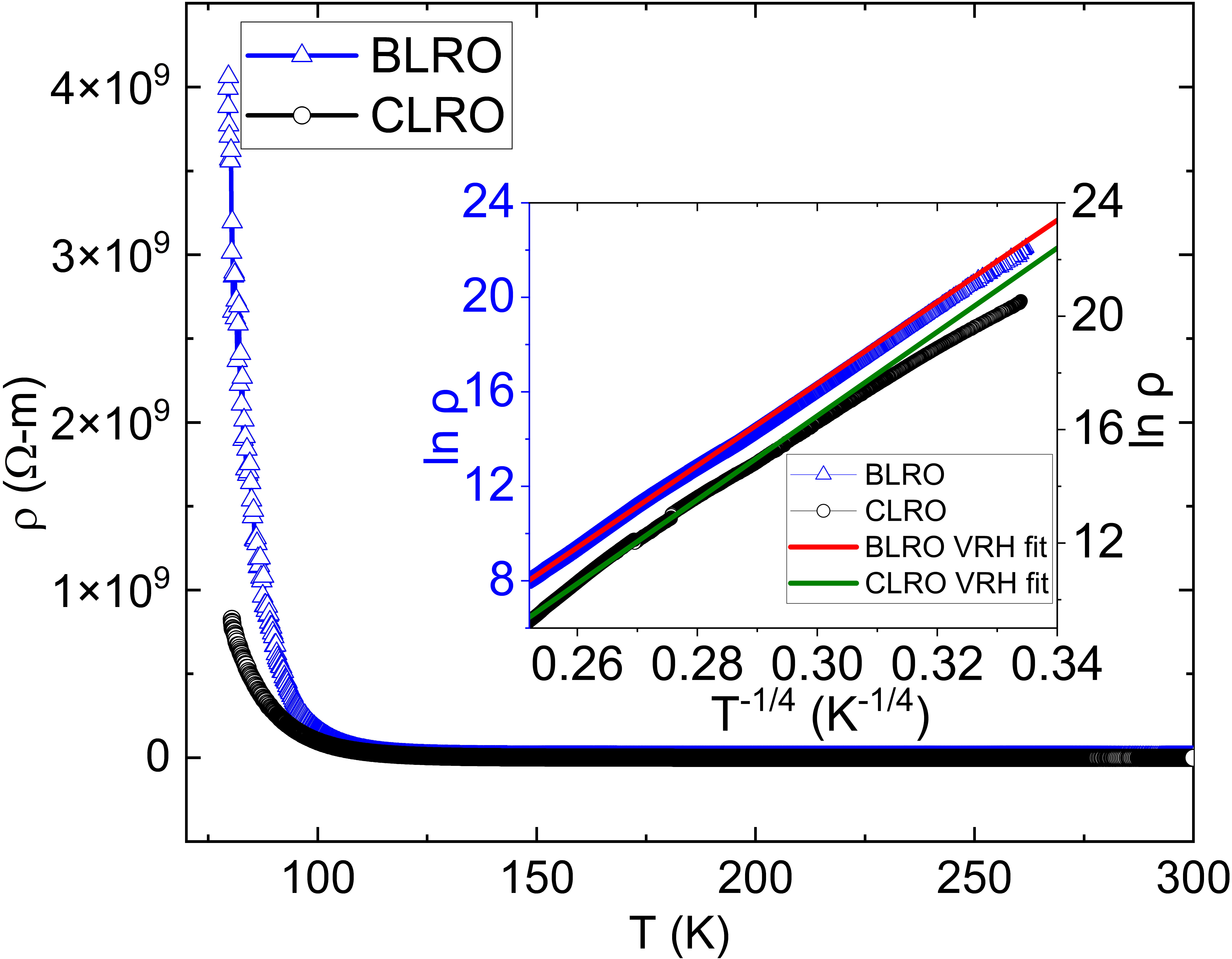}
         \caption{Temperature dependent resistivity measurements ($\rho$ $vs.$ $T$) of  BLRO and CLRO. Inset shows 3D variable range hopping (VRH) fit for the same.}
         \label{R_blro_clro}
         \end{figure}
Temperature  dependent resistivity measurements ($\rho$ $vs.$ $T$) of both BLRO and CLRO show an insulating behavior (FIG. \ref{R_blro_clro}). $\rho$($T$) does not follow an Arrhenius behavior except for a very narrow temperature region near 300 K (not shown here). Rather, resistivity of both systems follows a 3D Mott variable range hopping (VRH) behavior given by $\rho = \rho _{0} $exp$[({T_{0}/T})^{1/d+1}]$ (inset of FIG. \ref{R_blro_clro}), where $d$ = 3 is the dimensionality of the hopping. The fitted values of $T_0$  for BLRO and CLRO are 79.48 x 10$^{7}$ K and 47.75 x 10$^{7}$ K respectively. $\rho$($T$) of most polycrystalline Mott insulators with localized moments and defect states follow a similar 3D Mott VRH behavior \cite{3DVRH_Na2IrO3,3DVRH_Li2RhO4,3DVRH_DP_La2CoMnO6}.
\subsection{Specific Heat Capacity Measurements}
  \begin{figure}[]
         \centering
         \includegraphics[width=\columnwidth]{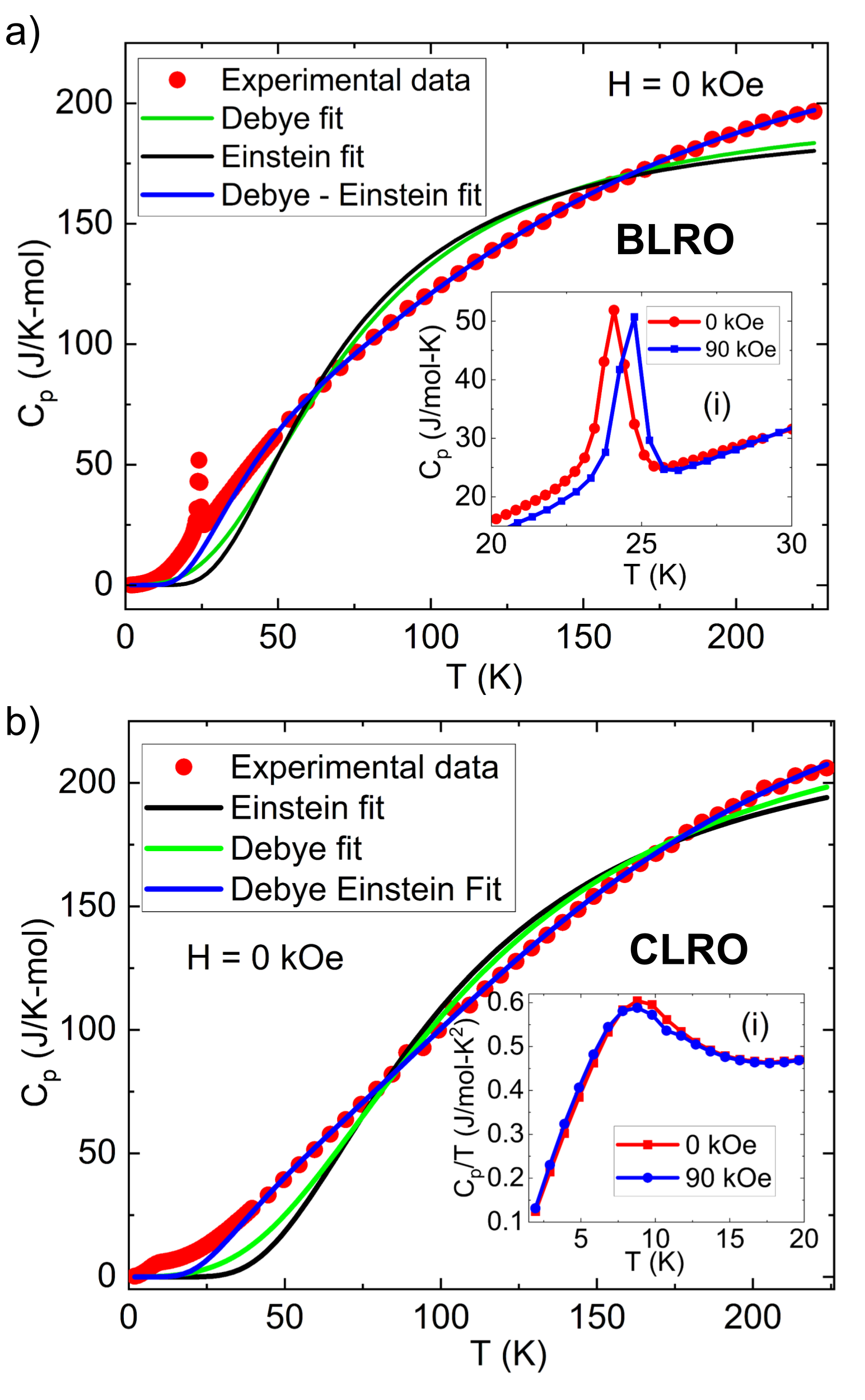}
         \caption{Temperature dependent heat capacity measurements of  a) BLRO and b) CLRO at $H$ = 0 kOe along with the Debye, Einstein and Debye - Einstein fit. Insets (i) display the temperature dependence of heat capacity at different fields.}
         \label{CT_u}
         \end{figure}
Further confirmation on magnetic phase transitions and Ru$^{5+}$ spin/pseudospin state of the synthesized A$_2$LaRuO$_6$ systems were obtained from temperature dependent specific heat measurements ($C_p$ $vs.$ $T$). FIG. \ref{CT_u}a shows a  clear $\lambda$-like anomaly for BLRO in $C_p$ at $T$ = 25 K indicative of the long-range AFM interaction and is consistent with the $T_N$ estimated from the magnetization measurements. A shift of the anomaly towards higher temperature by 90 kOe magnetic field (FIG. \ref{CT_u}a(i)) is counter intuitive to the conventional AFM system, rather pointing towards canted AFM ground state. No other anomaly is observed in $C_p$ for BLRO at $T$ = 7 K. This suggests that BLRO undergoes an AFM phase transition at 25 K and spin reorientation at $T$ = 7 K. On the other hand, a broader anomaly is observed around 9 K for CLRO, which shifts to a lower temperature when a 90 kOe magnetic field is applied (FIG. \ref{CT_u}b(i)). This implies a conventional AFM ground state with $T_N$ = 9 K for CLRO.

Magnetic specific heat ($C_{mag}$) for these materials is obtained by subtracting the lattice specific heat ($C_{lattice}$) from the total specific heat capacity ($C_p$). $C_{lattice}$ of a material follows either Debye model, Einstein model or combination of both. The Debye model accounts for the acoustic phonon modes, while the Einstein model describes the active optical phonon modes present in the system. The high temperature~$C_p$(T) of BLRO and CLRO does not fit either the Debye or Einstein model, as shown in the FIG. \ref{CT_u}. $C_{lattice}$  estimated from the Debye-Einstein model (Eq. \ref{DE_eqn}) fits the $C_p$ data very well in the temperature range 40 K - 225 K for both the BLRO and CLRO (FIG. \ref{CT_u}).
\begin{equation}
\centering
\label{DE_eqn}
  \begin{split}
     C_{lattice}(T) = n_{D}9R\left (  \frac{T}{T_{D}}\right )^{3}\int_{0}^{T_{D}/T}\frac{x^{4}e^{x}}{(e^{x}-1)^2} dx \\
    \\ +  n_{E}3R\left (  \frac{T_{E}}{T}\right )^{2}\frac{e^{T_E/T}}{(e^{T_E/T}-1)^2}
\end{split}
\end{equation}
The first and second term of Eq.\ref{DE_eqn} represent the Debye and Einstein model for specific heat capacity respectively. Here $n_{D}$, $n_{E}$, $T_E$, $T_D$, $R$, $x$ are Debye coefficent, Einstein coefficent, Einstein temperature, Debye temperature, universal gas constant and $\hbar\omega/k_BT$ respectively.\\ 
 \begin{table}[]
\caption{Parameters obtained from the Debye-Einstein fit.}
\vspace{5 mm}
\centering
\begin{tabular}{cccccc}
\hline
 && BLRO&&&CLRO \\ \hline
 $n_{E}$ && 4.11(7) &&& 2.8(1)\\
 $n_D$ && 5.75(4)&&& 8.69(9)\\
 $T_{E}$  && 130(2) K &&&150(5) K\\
 $T_D$  && 650(9) K&&& 690(13) K\\
 S$_{mag}$ && 10.82 J/K-mol&&& 11.35 J/K-mol\\
\hline
\end{tabular}
\label{Cpfit}
\label{DE_table}
\end{table} 
 \begin{figure}[]
         \centering
         \includegraphics[width=\columnwidth]{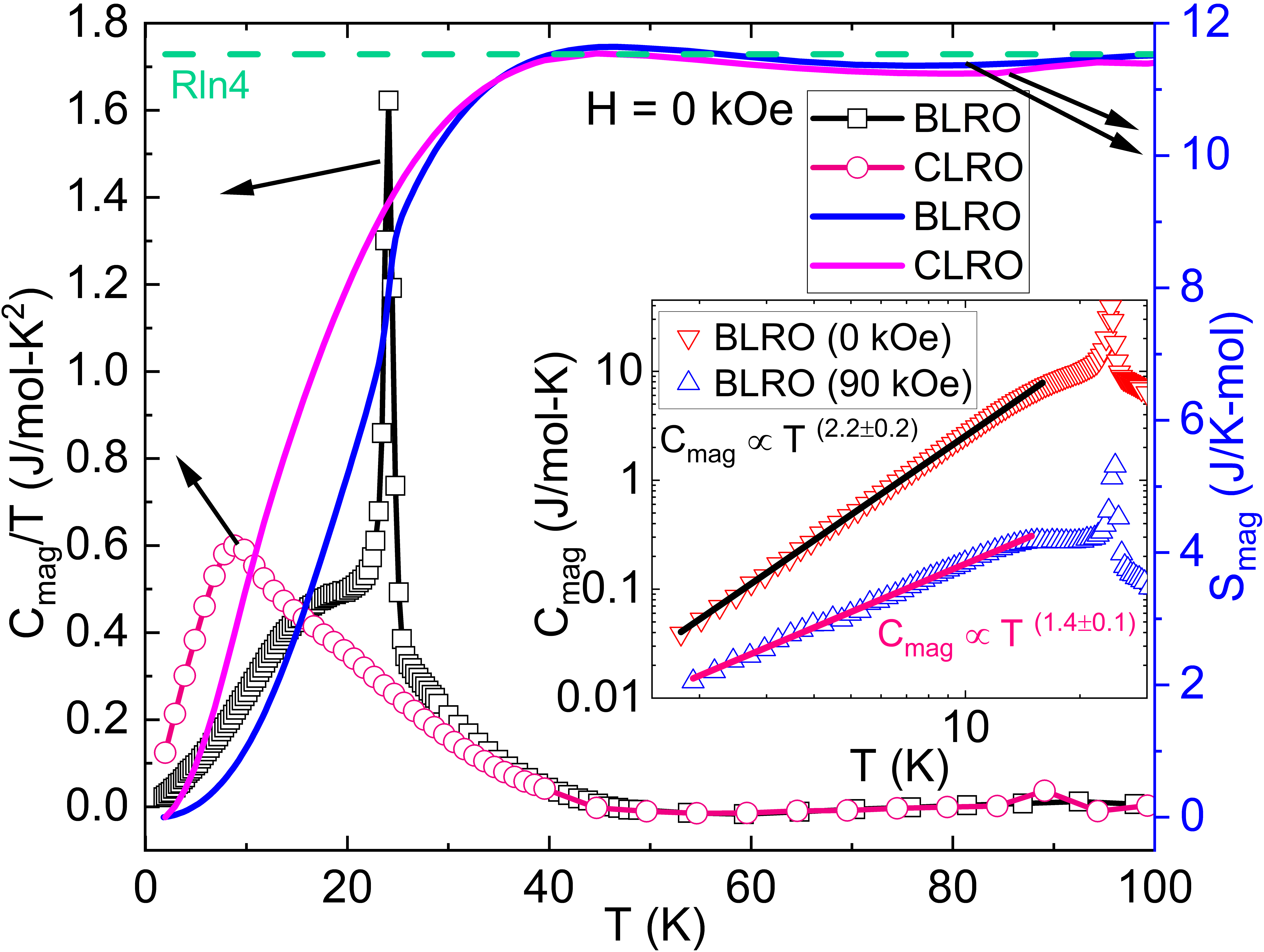}
         \caption{Temperature dependent Magnetic specific heat data (C$_{mag}$/$T$ $vs.$ $T$) is shown (left axis). Right axis displays the corresponding magnetic entropy (S$_{mag}$). Dashed green line represents the theoretically calculated S$_{mag}$, Rln4 (11.53 J/K-mol) for 4d$^3$ systems. Inset show a $T^{2.2}$ and $T^{1.4}$ dependence for the magnetic specific heat of BLRO at 0 kOe and 90 kOe respectively..}
         \label{S_blro_clro}
         \end{figure}

 For both BLRO and CLRO, $C_{mag}/T$ and the corresponding magnetic entropy ($S_{mag} = \int\frac{C_{mag}}{T}dT$) are shown in FIG. \ref{S_blro_clro}. A much broader transition is observed in $C_{mag}/T$ for CLRO compared to BLRO, possibly due to the partial spin freezing in CLRO above $T_N$. In both BLRO and CLRO, $S_{mag}$ reaches $Rln4$ near 50 K which is close to the theoretical value $Rln(2$S$+1)$ for $S = 3/2$. For both systems, $S_{mag}$ saturates at a temperature much lower than $|\Theta_{CW}|$ which may be due to the underestimation of $C_{lattice}$ at low temperature. TABLE \ref{Cpfit} provides the fitted parameters and the calculated $S_{mag}$ from the fit. The inset of FIG. \ref{S_blro_clro} shows a $T^{(2.2\pm0.2)}$ dependence of $C_{mag}$ for BLRO at 0 kOe below $T_N$, while a $T^{(1.4\pm0.1)}$ dependence is observed at 90 kOe. $C_{mag}$ of CLRO also follows a nearly $T^2$-dependence below $T_N$ (not shown here because of the small temperature range of fitting). Typically, the $T^2$-behavior in low temperature magnetic heat capacity is observed for gapless quantum spin liquids \cite{GSL1_Cm,GSL2_Cm}and in disordered Kitaev materials \cite{Li2RhO3_Cm,RudopedNa2IrO3_Cm}. Origin is usually connected with strong quantum fluctuations mediated by competing exchange interactions present in the magnetic sublattice of these systems. $T^2$-behavior is also observed in 2D Heisenberg antiferromagnets and other low-dimensional disordered antiferromagnetic systems\cite{Cm_LSMO,Cm_T2SG}. This behavior mostly originates from the quadratic spinwave dispersion relation.   Similar behavior was observed in antiferromagnetically ordered Pyrochlore Gd$_2$Sn$_2$O$_7$\cite{Cm_T2pyrochlore}. The near $T^2$-behavior of $C_m$ in long range antiferromagnetically ordered BLRO and CLRO is rather atypical in nature. It may originate from the special spinwave excitation present in these DPs which can be explained by low-dimensional magnetic exchange in the frustrated Ru-sublattice. 

\subsection{Electronic Structure Calculations}
\begin{figure}[]
\centering
\includegraphics[width=\columnwidth]{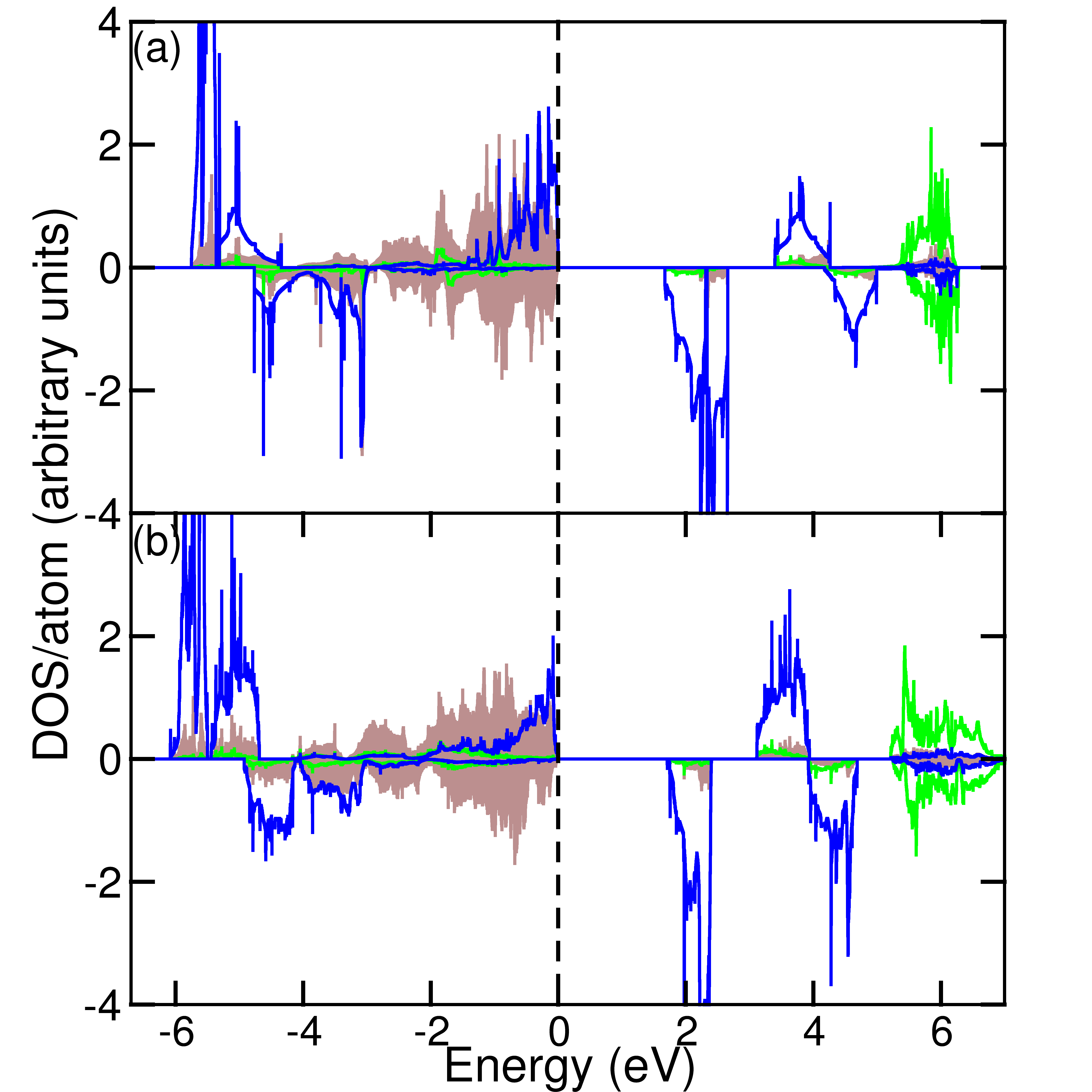}
\caption{The orbital projected DOS for (a) BLRO  and  (b) CLRO  with GGA+$U$(=4 eV). The blue, green, and brown curves represent the Ru-$4d$, La-$5d$, and O-$2p$ states respectively. The Fermi energy level has been set to zero in the energy scale.}
\label{DOS}
\end{figure}

The DFT based electronic structure calculations provide further insights into the materials from the microscopic point of view. We performed GGA+$U$ self-consistent calculations for both BLRO and CLRO, post-structural optimization. The Hubbard $U$ parameter was introduced due to the presence of a transition metal in the compound. The orbital projected GGA+$U$ ($U_{eff}$ = $U-J_H$ = 4 eV) spin-polarized density of states (DOS) for BLRO and CLRO are shown in FIG.\ref{DOS}(a) and (b) respectively. The most prominent feature that we found in the electronic structure is the presence of a large band gap at the E$_f$, which indicates the insulating nature of both double perovskites. Near the Fermi energy, the most dominant states are from the Ru-4$d$ states, as clearly visible in FIG.\ref{DOS}, which are broadly split into lower-lying $t_{2g}$ and upper-lying $e_{g}$ manifolds, due the octahedral crystal field splitting. The $t_{2g}$ manifold is completely filled in the majority spin channel and are located close to E$_f$. The minority spin channel is completely empty and the DOS lies at around 2 eV for both BLRO and CLRO. The completely empty e$_g$ states in both the spin channels are located far away from E$_f$ in the energy interval 3-5 eV in the conduction band. The empty 5$d$ states of La in both the spin channels, lie farther away in the energy scale and hence do not contribute actively to the electronic properties of the system. The O-2$p$ states are strongly hybridized with the Ru-4$d$ states as evident in FIG.\ref{DOS}, by the presence near the E$_f$. The gap due to exchange splitting between the $t_{2g}$ majority and minority spin channels is of the order of 1.6 eV and is marginally greater for CLRO due to the presence of structural distortions in the monoclinic structure. The Ru-4$d$ states located below 4eV from the E$_f$ in the valance band are the Ru anti-bonding states, which further extend upto 6 eV below the E$_f$.

The calculated values of the spin magnetic moment at the various sites are listed in Table \ref{spin_mmt}. The absence of substantial value of magnetic moment at the Ba, Ca, La, and O sites suggests that Ru is the only magnetically active site in BLRO and CLRO. The small value of the spin magnetic moment present at the O site is due to strong hybridization within the RuO$_6$ octahedra as previously visualized in the overlap of Ru-4$d$ and O-2$p$ DOS near the E$_f$. The total magnetic moment of the system is found to be 3 $\mu_B$/formula unit. From the combined findings of DOS and magnetic moments, we can conclude that Ba/Ca and La are in inert configurations with a nominal valence state of +2 and +3, respectively, and O is in -2 [$2p^6$] electronic state. On the other hand, Ru is +5 [$4d^3$] electronic state, with a high spin state of S= $\frac{3}{2}$. The above findings are consistent with the magnetization measurements discussed previously. Furthermore, calculations were performed with the range of values of $U$ and we find that there are no major alterations in the electronic and magnetic properties of BLRO and CLRO with the variation of $U$. Interestingly, the electronic structure calculation in the absence of Hubbard $U$, also shows a gap in the DOS, at the Fermi level. Since the above-mentioned gap at the Fermi level is due to the exchange splitting of the $t_{2g}$ majority and minority spin bands, the BLRO and CLRO clearly fall under the category of correlation-driven insulators.
\begin{figure}[]
\centering
\includegraphics[width=\columnwidth]{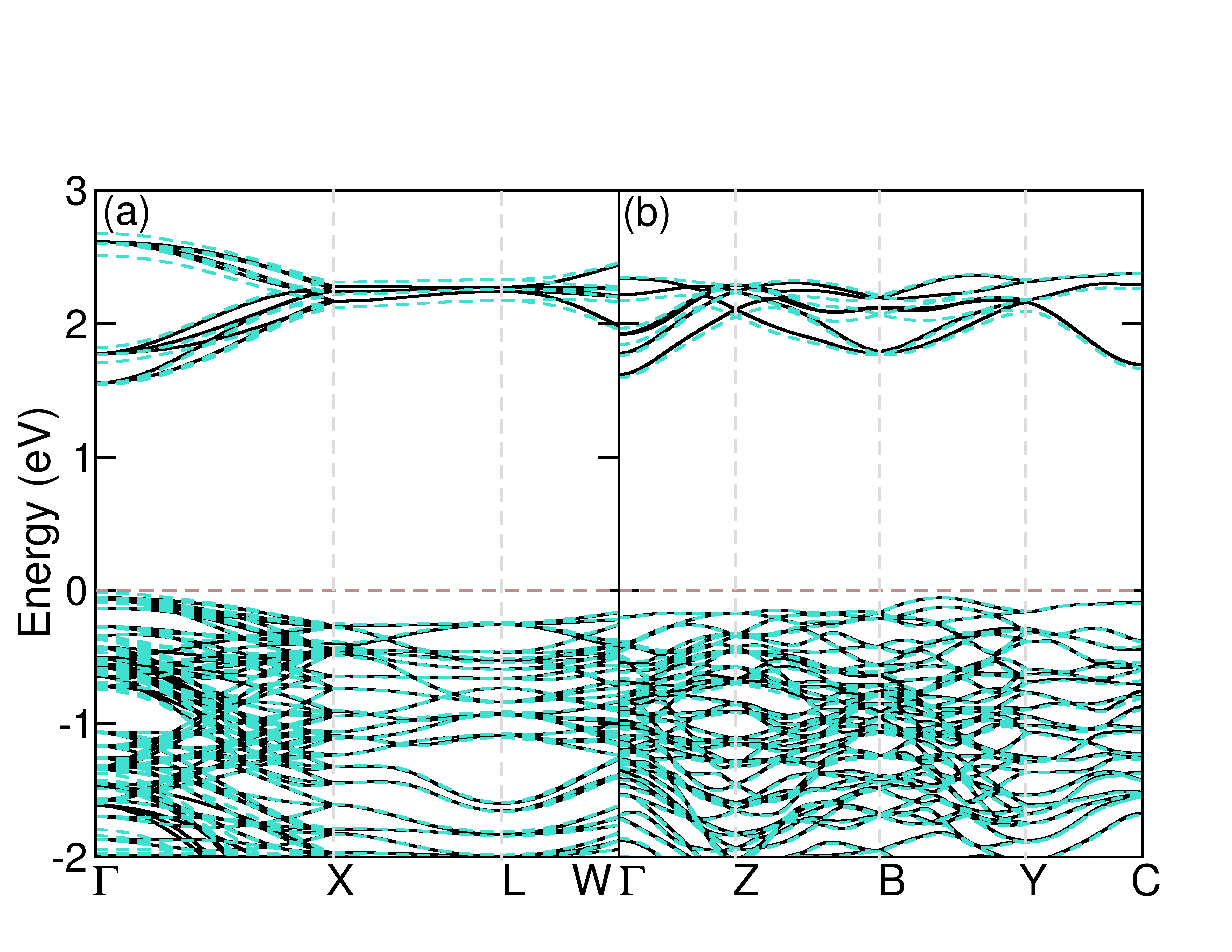}
\caption{The calculated band-structure for (a) BLRO  and  (b) CLRO. The black and turquoise dotted curves represent the GGA+$U$ and GGA+$U$+SOC bandstructures respectively. The Fermi energy level has been set to zero in the energy scale.}
\label{BAND}
\end{figure}
\begin{table}[]
\caption{ The GGA+$U$ calculated values of the spin magnetic moment for BLRO and CLRO. The spin magnetic moments are mentioned in units of $\mu_B$.}
\vspace{5 mm}
\centering
\begin{tabular}{cccccc}\hline
 Site && BLRO &&& CLRO \\ \hline
 Ba/Ca && 0.01 &&& 0.06\\
 La && 0.01 &&& 0.03\\
 Ru  && 2.36 &&& 2.35 \\
 O  && 0.07 &&& 0.1 \\
\hline
\end{tabular}
\label{spin_mmt}
\end{table} 
Experimental measurements clearly show that in both BLRO and CLRO compounds, AFM correlations are dominating in nature, with the possibility of low-temperature, short-range FM correlation with spin canting. The $\frac{1}{\chi}-T$ measurements show the CW temperature ($\Theta$) for BLRO is almost twice that of CLRO, indicating the magnetic exchange interactions are stronger in BLRO than CLRO. Therefore, to understand the 
underlying magnetism, we have done the magnetic exchange interaction calculations for both BLRO and CLRO. We have considered the nearest neighbor exchange path between the different Ru ions to calculate the hopping interaction. The calculated magnetic exchange (J) interactions incorporating the SOC effect are found to be AFM in nature. The computed values are 10.92 meV and 5.93 meV for BLRO and CLRO 
respectively. The calculated values of the nearest neighbor exchange interaction are in close
agreement with experimental findings and the ratio of the $J$'s for BLRO and CLRO ($\approx$ 1.8), is in good agreement with the experimentally measured ratio of the CW temperatures 
($\Theta$) of the BLRO and CLRO. Moreover, the nearest neighbor strong AFM exchange 
interactions in the cubic structured BLRO induces strong magnetic frustration in the 
tetrahedrally coordinated Ru sublattice. Therefore, in the low-temperature regime, due to the 
enhancement of the quantum fluctuations, the short-range interactions become proactive and 
finally lead to the spin canting. The observed minute feature in the magnetic measurement of the BLRO close to 8 K, may arise due to the possible spin non-collinearity in the system. 
\begin{figure}[]
\centering
\includegraphics[width=1.1\columnwidth]{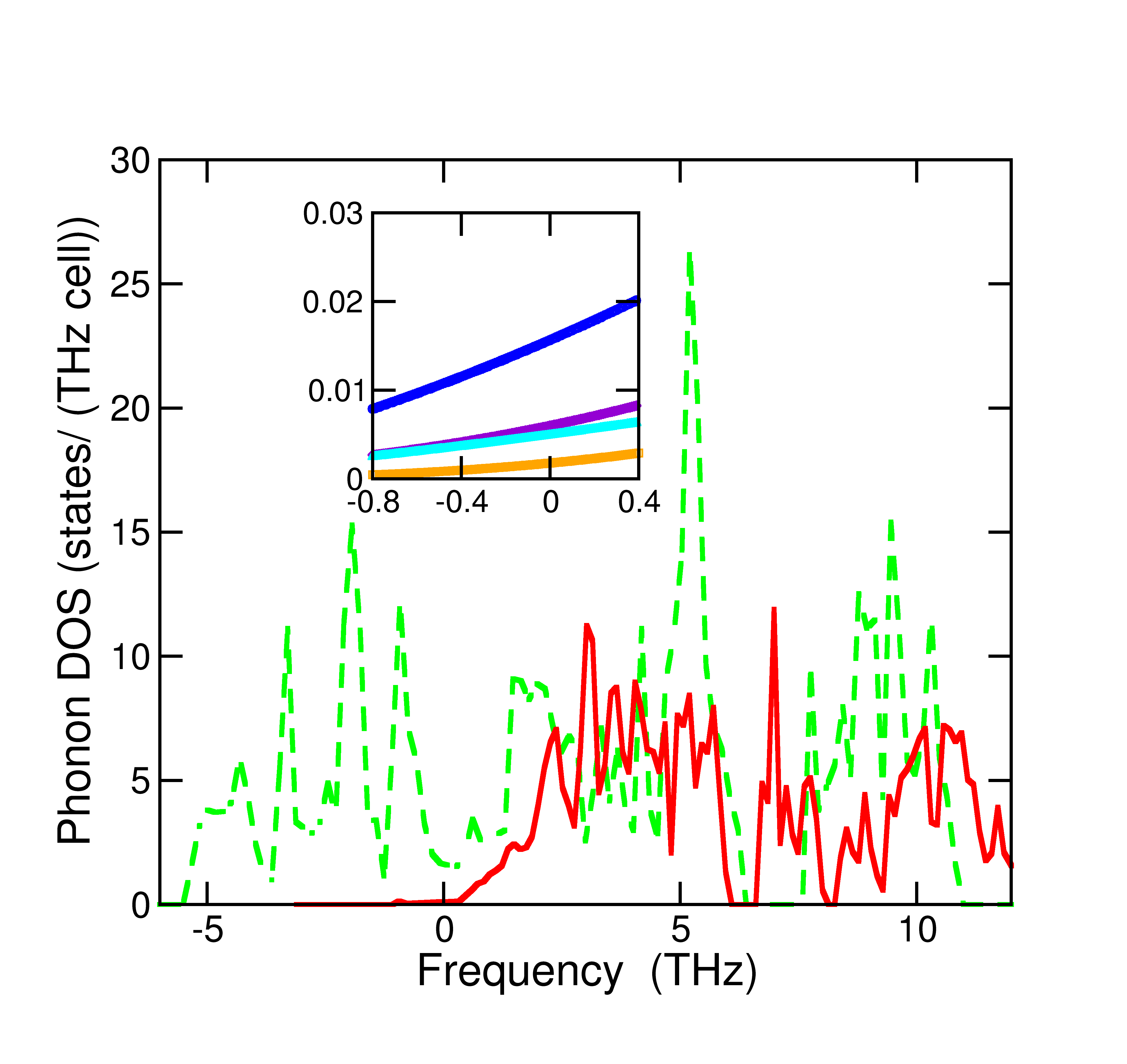}
\caption{The calculated phonon density of states for Sr$_2$LaRuO$_6$. The red (continuous) and green (dotted) lines represent the phonon DOS considering monoclinic and cubic symmetries respectively. The inset shows the zoomed version of the atom site projected phonon densities for the monoclinic symmetry, where the blue, orange, violet, and cyan curves represent the contribution from Sr, Ru, La, and O atoms respectively.}
\label{PHONON}
\end{figure}

In FIG. \ref{BAND}, the band structure for BLRO and CLRO are shown along the high symmetry $K$-points of the respective space groups. We see that the band structures have a very similar band dispersion, both in the presence and absence of the incorporation of SOC in the calculations. Except for the breaking of degeneracy-driven changes in the band dispersion, the SOC does not play a major role in deriving the underlying electronic structure of these compounds. Furthermore, if we look into the magnetic moment at the Ru site for GGA+$U$+SOC along the [001] direction, the spin magnetic moments for BLRO and CLRO show negligible changes with a value of 2.34  $\mu_B$. The orbital magnetic moment is calculated to be 0.01 $\mu_B$, which is also very small. From the above findings, we can conclude that the impact of SOC is limited in both compounds which can be attributed to the half-filled 4$d$ state of Ru [$t_{2g}^3$], which leads to quenching of the orbital degrees of freedom resulting in a negligible value of the orbital magnetic moment. Thus, even in the presence of 4$d$ magnetic element Ru, both the compounds fall under the description of weak SOC regime and the insulating nature can be classified in the conventional Mott regime. 
\begin{figure}[]
\centering
\includegraphics[width=\columnwidth]{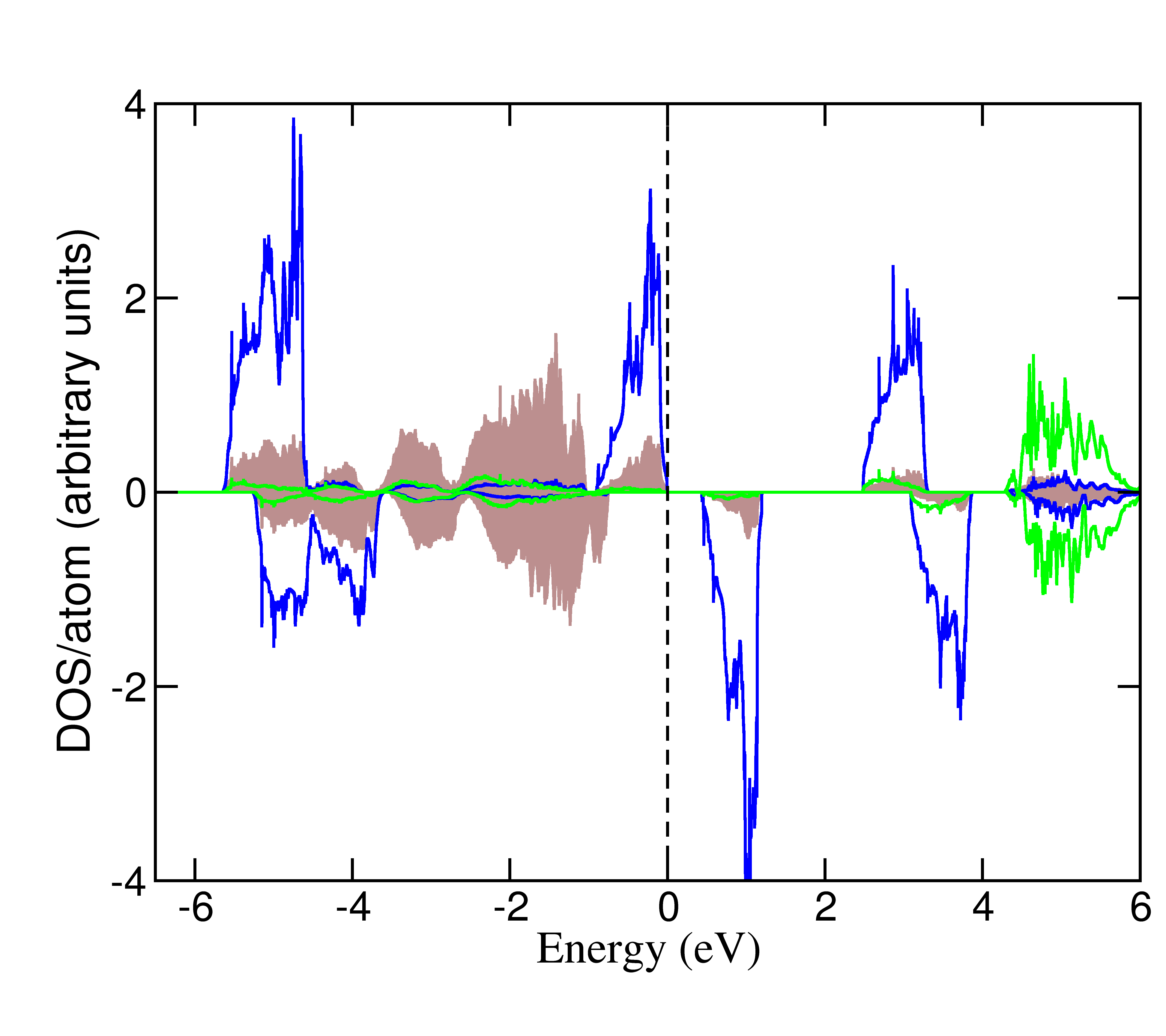}
\caption{The orbital projected DOS for CaSrLaRuO$_6$ with GGA+$U$(=4 eV). The blue, green, and cyan curves represent the Ru-$4d$, La-$5d$, and O-$2p$ states respectively. The Fermi energy level has been set to zero in the energy scale.}
\label{Sr_DOS}
\end{figure}

Both theoretically and experimentally, the Ba and the Ca variants have been investigated thoroughly, 
and the obtained results are consistent with each other. Thus, it is a natural trend to explore the 
possible Sr variant of this series. However, the synthesis of the Sr
variant in the double-perovskite form is not possible from experiments. We also tried to explore the possible reason behind the unsuccessful attempt for the synthesis of the Sr variant, via phonon investigation. We have carried out the calculations of the Sr$_2$LaRuO$_6$ to test the dynamical stability via phonon density of states as shown in FIG.\ref{PHONON}. We found that the Sr variant is very unlikely to be in the cubic symmetry phase as large negative frequency phonon modes existed in the phonon DOS 
(green dotted curve), as shown in the FIG.\ref{PHONON} . However, in the case of the monoclinic symmetry structure of the Sr$_2$LaRuO$_6$,  the dynamical instability has been comparatively reduced as evident in the phonon DOS (red curve). We have found from the atom projected phonon DOS (inset of FIG.\ref{PHONON}), that the negative frequency contributions are predominantly arising from the Sr states, which is very much intuitive. We have also investigated the 50$\%$ Sr doped in the Ca$_2$LaRuO$_6$ and compared the results with the experiments. The calculated DOS for CaSrLaRuO$_6$ is shown in FIG.\ref{Sr_DOS}. We found that the main features of the electronic structure remain intact, including the charge and the spin states. The magnetic moment (2.01 $\mu_B$/Ru site) is also marginally changed, which is consistent with the experimental observation. The prominent difference that distinguishes this from the Ca variant is that the band gap has decreased in the case of the 50$\%$ doped case. We also calculated the magnetic exchange (5.91 meV), which is marginally reduced compared to that of the undoped Ca compounds, which also has been seen in the experiments. 

\section{Conclusions}
Detailed structural analysis by PXRD investigation and electronic structural calculation suggest that in the A$_2$LaRuO$_6$ (Ca-Ba) series, maximum unit-cell volume with the largest atom at the A-site stabilizes the ordered DP structure. The unit cell undergoes monoclinic distortion with decreasing size. This is evident from the nonexistence of Sr-doped BLRO and realization of monoclinic crystal structure in low Sr-doped CLRO. Although it is really intriguing that in this series Sr$_2$LaRuO$_6$ and 75\% Sr-doped CLRO do not form in any of the crystal structures, whereas the phonon DOS calculation does not rule out the formation of SLRO in a monoclinic structure. 

 The change in symmetry also affects the magnetic properties of the material. BLRO possesses an FCC Ru unit cell having Ru-tetrahedron with a Ru-Ru distance of 6.04061 (12)$\AA$. This results in a geometrically frustrated magnetic sublattice which is clearly reflected in a very high frustration parameter (=12.45). An earlier neutron diffraction study in BLRO with a monoclinic crystal structure suggested collinear Type IIIa magnetic ordering in FCC pseudocubic sublattice at 2K\cite{battle1983crystal}. But our magnetization data showed signatures of spin canting below 10 K and a clear cubic crystal structure without any structural transition. Further neutron diffraction study is required in this system to identify the ground-state magnetic structure. However, because of the monoclinic distortion in CLRO, the Ru-sublattice forms Ru isoscale triangles. It has a frustration index comparable to that of BLRO, mostly because of competing magnetic interactions. Previous neutron diffraction measurements in CLRO suggested a Type I magnetic ordering in fcc pseudocubic sublattice which can be explained from our magnetization data \cite{battle1983crystal}. The $T^2$-bahavior of the low-temperature magnetic heat capacity indicates low-dimensional magnetic exchange and magnetic frustration in BLRO and CLRO. It is worth to note that inelastic neutron measuremenet in Ba$_2$YRuO$_6$ system confirmed the gapped spinwave exciation with 5 meV spin gap \cite{magmoment_2}. The spinwave exciations in BLRO and CLRO can be very different from its yttrium counterparts, which can only be confirmed from further millikelvin heat capacity measurement and ineleastic neutron scattering measurement.
 
 $\mu_{eff}$ and $S_{mag}$ calculated from the magnetization and heat capacity measurements imply a negligible effect of SOC in these systems, the same is corroborated by the DFT calculations. This points towards a conventional L-S coupling in this 4d$^3$ system, which defers from the theoretical proposed coupling limit for 4d$^3$/5d$^3$ systems.\cite{theoretical4d35d3}. A clear understanding of the electronic structure of these systems can be investigated from the resonant inelastic X-ray scattering (RIXS) measurements. 
 
 In summary, our work proves that BLRO and CLRO have a non-relativistic Mott insulating AFM ground state with interesting spinwave dispersion. However, SLRO is situated near a structural instability, making it impossible to synthesize in a single-phase.
\begin{acknowledgments}
AAA and SM thank Prof. Arumugam Thamizhavel and the TIFR, Mumbai, for magnetization and heat capacity measurement facilities. AAA thanks the CIF, IIT Palakkad for the experimental facilities. SM thanks the DST INSPIRE Faculty grant for research funding. 
\end{acknowledgments}
\appendix*
\section{Ca$_{2-x}$Sr$_{x}$LaRuO$_{6}$ (x = 0.5, 1)}{\label{appendix}}

\begin{figure}[]
\centering
\includegraphics[width=\columnwidth]{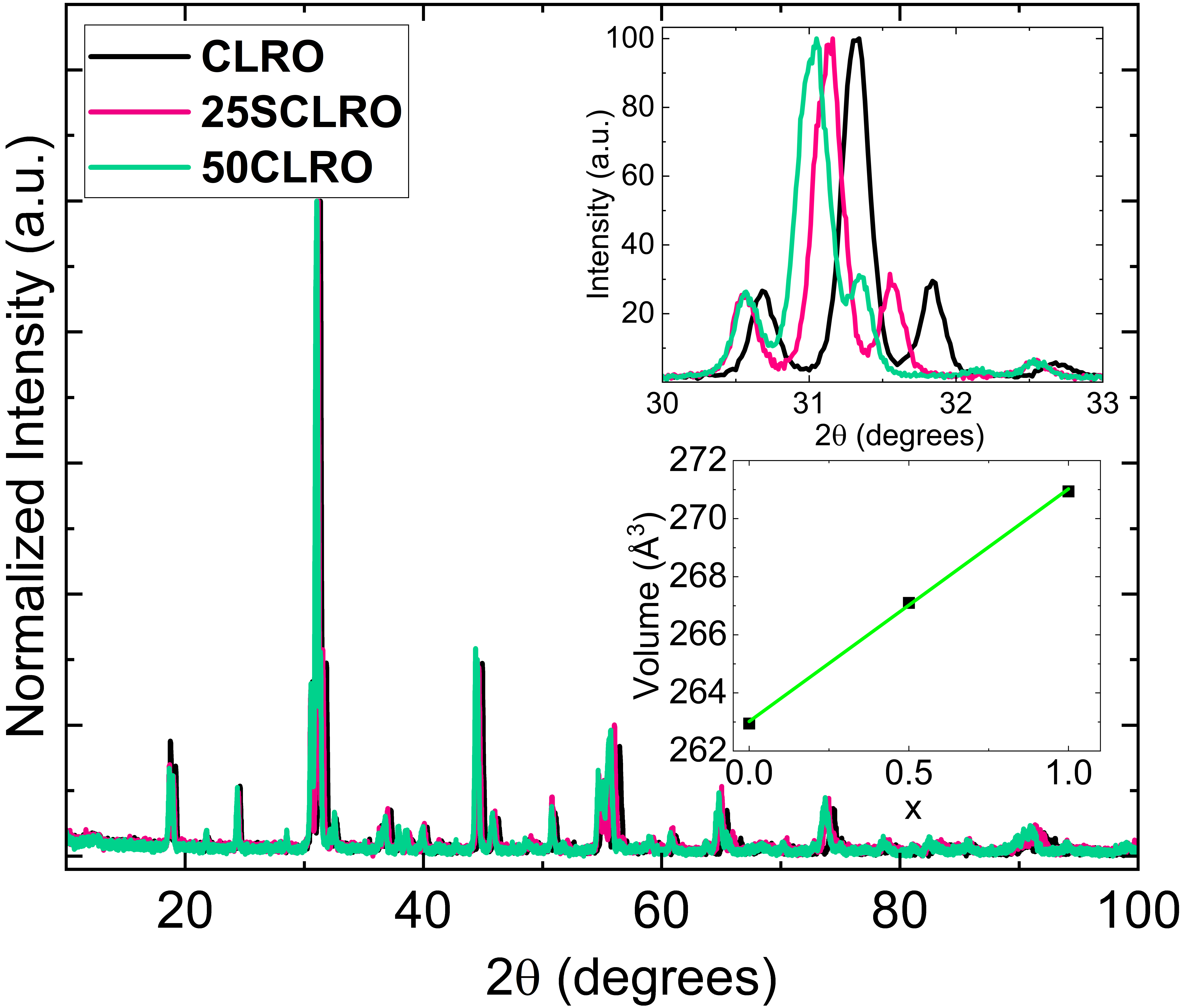}
\caption{PXRD data of Ca$_{2-x}$Sr$_{x}$LaRuO$_{6}$ (x = 0, 0.5, 1).Inset (i) shows the zoomed plot of the shifted peaks. Inset (ii) displays the variation of unit cell volume with change in the doping fraction (x).}
\label{clro_u_d}
\end{figure}

PXRD data of the single-phase polycrystalline Sr-doped CLRO systems clearly show a shift in the peak position (inset (i) of FIG. \ref{clro_u_d}). Here Ca$_{1.5}$Sr$_{0.5}$LaRuO$_{6}$ and CaSrLaRuO$_{6}$ are represented by 25SCLRO and 50SCLRO, respectively. The PXRD patterns of Sr-doped CLRO is fitted with \textit{P2$_1$/n} space group. Sr-doping in CLRO results in a increase in all the lattice parameters (TABLE \ref{rrdoped}) and a corresponding increase in the volume of the unit cell. The linear increase in unit cell volume upon Sr-doping of CLRO follows the Vegard's law as shown in inset (ii) of FIG. \ref{clro_u_d}.
\begin{table}[]
\caption{Refined lattice parameters of Sr-doped CLRO system using \textit{P2$_1$/n} space group: Ca$_{1-x}$Sr$_{x}$LaRuO$_{6}$.}
\vspace{5 mm}
\centering
\begin{tabular}{ccccc}
\hline
x   & a (Å)     & b (Å)        & c (Å)        & $\beta$ ($^\circ$)        \\ \hline
0   & 5.6127(1) & 5.816(9)  & 8.0538(2) & 90.224(3) \\ 
0.5 & 5.6557(1) & 5.8362(1) & 8.0921(2) & 90.253(2) \\ 
1   & 5.7014(2) & 5.84(1)   & 8.1293(3) & 90.298(3) \\ \hline
\end{tabular}
\label{rrdoped}
\end{table}

\begin{figure}[]
\centering
\includegraphics[width=\columnwidth]{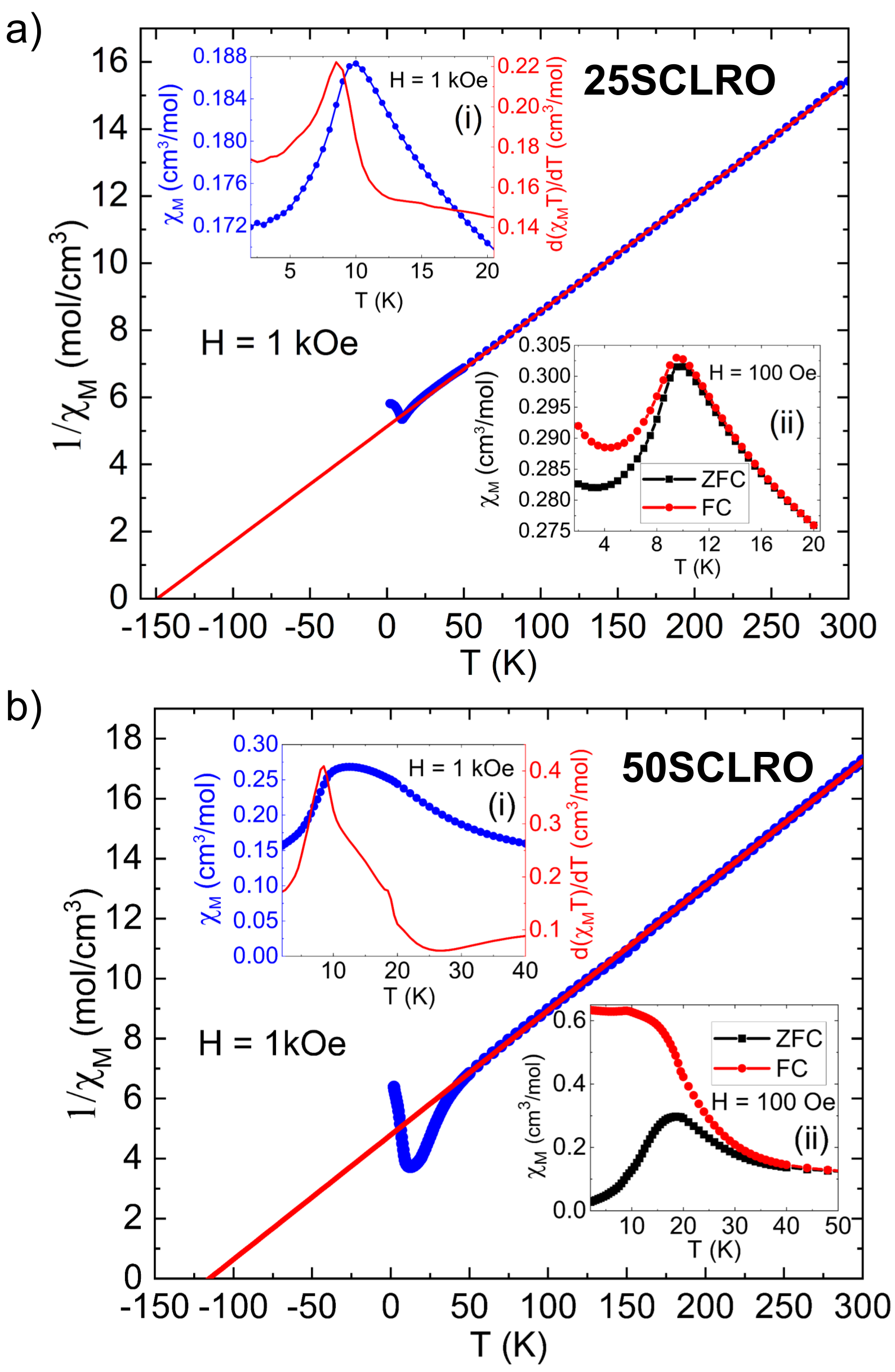}
\caption{Temperature dependent inverse magnetic susceptibility data for Ca$_{2-x}$Sr$_{x}$LaRuO$_{6}$ a) x = 0.5 and b) x =1 at an external magnetic field, H = 1 kOe with the CW fit (red line) from 50 K - 300 K. Inset (i) shows the molar susceptibility ($\chi_M$) and $\frac{\mathrm{d}(\chi_{m}T)}{\mathrm{d} T}$ $vs.$ $T$ and inset (ii) shows the Zero Field Cooled (ZFC) and Field Cooled (FC) $\chi_M$ $vs.$ $T$ at low temperature near magnetic phase transitions for both the systems in their respective panels.}
\label{MT_d}
\end{figure}
\begin{table}[]
\caption{$T_N$, $\Theta$ and $\mu_{eff}$ obtained from the high temperature CW fit of Ca$_{2-x}$Sr$_{x}$LaRuO$_{6}$.}  
\vspace{5 mm}
\centering
\begin{tabular}{ccccc}
\hline
$x$& $T_N$ (K) &$\Theta$ (K) &$f$ &$\mu_{eff}$ ($\mu_B$)\\ \hline
0& 8.86&  -137.5(3)&15.52&4.004(4)\\
0.5& 8.49& -149.3(3)&17.5&4.318(3)\\
1& 8.53& -115.4(4)&13.52&3.926(3)\\
\hline
\end{tabular}
\label{MT_utable}
\end{table}

1/$\chi_M$ $vs.$ $T$ plots of 25SCLRO and 50SCLRO are shown in FIG. \ref{MT_d} a and b, respectively. Both of them follow a CW behavior in the high-temperature regime with strong AFM interactions with a slight decrease in $\Theta$ with Sr-doping. The AFM transitions in 25SCLRO and 50SCLRO are shown in the insets (i) of FIG. \ref{MT_d} by the prominent peak in the respective  $\frac{\mathrm{d}(\chi_{M}T)}{\mathrm{d} T}$ $vs.$ $T$ and peak drop in $\chi_M$ $vs.$ $T$ data. The ZFC -FC plot of 50SCLRO (FIG. \ref{MT_d}b(ii)) clearly shows a large separation in the $\chi_M$ $vs.$ $T$ plot at low field, indicating a possible spin-glass state. A linear $M vs. H$ behavior further confirmed the AFM ground state of these doped systems (not shown here). Magnetic parameters are given in TABLE \ref{MT_utable}.
\providecommand{\noopsort}[1]{}\providecommand{\singleletter}[1]{#1}%

\end{document}